\documentclass{aa}  
\usepackage{graphicx}
\usepackage{txfonts}
\usepackage{xcolor}
\usepackage{makecell}
\usepackage{booktabs}

\usepackage{multirow}
\usepackage{scalerel}
\usepackage{tikz}
\usetikzlibrary{svg.path}

\definecolor{orcidlogocol}{HTML}{A6CE39}
\tikzset{
  orcidlogo/.pic={
    \fill[orcidlogocol] svg{M256,128c0,70.7-57.3,128-128,128C57.3,256,0,198.7,0,128C0,57.3,57.3,0,128,0C198.7,0,256,57.3,256,128z};
    \fill[white] svg{M86.3,186.2H70.9V79.1h15.4v48.4V186.2z}
                 svg{M108.9,79.1h41.6c39.6,0,57,28.3,57,53.6c0,27.5-21.5,53.6-56.8,53.6h-41.8V79.1z M124.3,172.4h24.5c34.9,0,42.9-26.5,42.9-39.7c0-21.5-13.7-39.7-43.7-39.7h-23.7V172.4z}
                 svg{M88.7,56.8c0,5.5-4.5,10.1-10.1,10.1c-5.6,0-10.1-4.6-10.1-10.1c0-5.6,4.5-10.1,10.1-10.1C84.2,46.7,88.7,51.3,88.7,56.8z};
  }
}

\newcommand\orcidicon[1]{\href{https://orcid.org/#1}{\mbox{\scalerel*{
\begin{tikzpicture}[yscale=-1,transform shape]
\pic{orcidlogo};
\end{tikzpicture}
}{|}}}}

\usepackage{hyperref}
\begin{document} 
\title{Application of Convolutional Neural Networks to time domain astrophysics. 2D image analysis of OGLE light curves}

\author{N. Monsalves\inst{1,2\orcidicon{0000-0002-4129-8195}}, M. Jaque Arancibia\inst{1\orcidicon{0000-0002-8086-5746}}, A. Bayo\inst{2\orcidicon{0000-0001-7868-7031}}, P. S\'anchez-S\'aez\inst{2\orcidicon{0000-0003-0820-4692}}, R. Angeloni\inst{3\orcidicon{0000-0001-7978-7077}}, G. Damke\inst{4\orcidicon{0000-0002-2651-7038}}, J. Segura Van de Perre\inst{1\orcidicon{0009-0002-9040-8263}}}
\institute{$^{1}$Departamento de Astronom\'ia, Universidad de La Serena, Avenida Juan Cisternas 1200, La Serena, Chile \\
$^{2}$European Southern Observatory, Karl-Schwarzschild-Str. 2, D-85748 Garching, Germany \\
$^{3}$Gemini Observatory, NSF's NOIRLab, Av. J. Cisternas 1500 N, 1720236 La Serena, Chile\\
$^{4}$ Cerro Tololo Inter-American Observatory/NSF's NOIRLab, Casilla 603, La Serena, Chile\\
\email{nicolas.monsalves@userena.cl}
}
\date{}

  \abstract
   {In recent years the amount of publicly available astronomical data has increased exponentially, with a remarkable example being large scale multiepoch photometric surveys. This wealth of data poses challenges to the classical methodologies commonly employed in the study of variable objects. As a response, deep learning techniques are increasingly being explored to effectively classify, analyze, and interpret these large datasets. In this paper we use two-dimensional histograms to represent Optical Gravitational Lensing Experiment (OGLE) phasefolded light curves as images. We use a Convolutional Neural Network (CNN) to classify variable objects within eight different categories (from now on labels): Classical Cepheid (CEP), RR Lyrae (RR), Long Period Variable (LPV), Miras (M), Ellipsoidal Binary (ELL), Delta Scuti (DST), Eclipsing Binary (E), and spurious class with Incorrect Periods (Rndm). We set up different training sets to train the same CNN architecture in order to characterize the impact of the training. The training sets were built from the same source of labels but different filters and balancing techniques were applied. Namely:  Undersampling (U), Data Augmentation (DA), and Batch Balancing (BB). The best performance was achieved with the BB approach and a training sample size of $\sim$370000 stars. Regarding computational performance, the image representation production rate is of $\sim$76 images per core per second, and the time to predict is $\sim$ 60$\, \mu\text{s}$ per star. The accuracy of the classification improves from $\sim$ 92\%, when based only on the CNN, to $\sim$ 98\% when the results of the CNN are combined with the period and amplitude features in a two step approach. This methodology achieves comparable results with previous studies but with two main advantages: the identification of miscalculated periods and the improvement in computational time cost.}

\titlerunning{Image representation classification of OGLE LCs}
\authorrunning{N. Monsalves et al.}

\maketitle

%
\nolinenumbers

\section{Introduction}
By definition, variable stars exhibit detectable changes in brightness over time. These changes can result from physical processes intrinsic to the stars or geometric processes that affect them. Geometric processes include eclipses due to a companion \citep{2000ApJ...529L..45C} or rotation \citep{2005LRSP....2....8B}. The intrinsic processes include pulsation, flares \citep{2013ApJS..209....5S}, and cataclysmic eruptions \citep{2001cvs..book.....H}. \newline
Time domain astronomy has played a fundamental role in our current knowledge of the Universe \citep{1929PNAS...15..168H, 2016ApJ...826...56R, 2019ApJ...876...85R}. Historically, these objects have been a great source of information because they allow us to determine fundamental astrophysical relationships from the nature of variation. One excellent example is the period-luminosity relation discovered by \cite{1912HarCi.173....1L}, which was fundamental to calculate the distance to M31 \citep{1925PA.....33..252H} thus marking the beginning of extragalactic astronomy \citep{schneider2006extragalactic}. \newline
Based on their lightcurve (LC) morphology, variable stars can be broadly classified into regular, semiregular and irregular variables. Regular variables exhibit a clear pattern that repeats over time, whereas irregular variables show no obvious signs of periodicity. Semiregular variables display some signs of periodicity, along with stochastic variations \citep{2007uvs..book.....P}. Aspects of the morphology of different types of LCs can be quantified in so-called features that vary in complexity regarding their computation, from simple amplitudes to more involved accounting of the timescales of variations. Such features encode hints on the structure and evolutionary state of these variable stars \citep{catelan2015pulsating}.\newline
In this context, regular variables are particularly useful since the characteristics of their LCs allow us to infer fundamental astrophysical parameters. The period-luminosity relations of pulsating stars, including Cepheids, RR Lyrae, and Mira, play a crucial role in determining the cosmic distance ladder across the universe \citep{2011ApJ...730..119R,2018MNRAS.480.4138M,2023MNRAS.523.2369S}. Eclipsing systems, on the other hand, allow us to directly estimate mass, radius, temperature, and absolute luminosity of the system's components. These parameters are necessary to test the models of stellar structure and evolution \citep{2010A&ARv..18...67T}. They are the primary source of empirical information about the properties of stars, making them a foundational pillar of modern astrophysics \citep{2012ocpd.conf...51S}. \newline

In the last decades, multiepoch photometric surveys have been developed to address different scientific problems. These surveys obtain time series of different objects with an extended temporal coverage, such as: Massive Compact Halo Objects (MACHO) \citep{1997ApJ...486..697A}, The All Sky Automated Survey (ASAS) \citep{2002AcA....52..397P},The Optical Gravitational Lensing Experiment (OGLE) \citep{2003AcA....53..291U}, The Northern Sky Variability Survey (NSVS) \citep{2004AJ....127.2436W}, The Catalina Real-Time Transient Survey (CRTS) \citep{2014ApJS..213....9D}, The Zwicky Transient Facility (ZTF) \citep{2019PASP..131a8002B} and the \textit{Gaia} mission \citep{2023A&A...674A...1G}, just to cite a few. These surveys have generated (or continue to do so) large amounts of data, pushing towards multidisciplinary efforts to develop new methodologies to accurately and efficiently analyze them. This scenario of big data will only increase with the imminent arrival of new telescopes such as the \textit{Vera C. Rubin} Observatory and its Legacy Survey of Space and Time \citep{2019ApJ...873..111I}, which will revolutionize the way astronomy works, generating tens of terabytes of data each night \citep{2019ApJ...873..111I}. \newline
A myriad of automatic classification methodologies have surged in the community to tackle this challenge. Examples of such methodologies can include conventional supervised machine learning techniques operating over a space of features that characterize the LCs, and subsequently apply these features to the classification of the LC \citep{2012MNRAS.427.1284P, 2015arXiv150600010N, 2021AJ....161..141S}. These methodologies have shown robust results. However, feature extraction is a complex process that typically requires significant time and research to be conducted effectively \citep{2019MNRAS.482.5078A, 2021AJ....161..141S}.\newline
In the last few years deep learning methods have achieved highly accurate results \citep{2018NatAs...2..151N,2020MNRAS.493.2981B,2022AJ....164..263M}. These methods automatically extract important features through a trial and error training process. They require large amounts of data for training and arise as a natural response to the current situation of big data. The great popularity of these methods in recent years is due to the current availability of more powerful computational resources, such as Graphics Processing Unit (GPU) \citep{ZHANG2018146}.\newline
The most common representation of variable stars includes sequential data, statistical / characteristics abstraction of the former, or a combination of both. Sequential data consist of time series or different Adjustments of the time series. For example in \cite{2020MNRAS.493.2981B} they used the difference with previous measurements in time and magnitude. The tabular data can consist of features describing the LC, either from statistical information or astrophysical knowledge \citep{2015arXiv150600010N}. Another less explored way of representing LCs is by using an image. \citet{Mahabal_2017} use the raw LC with a 2D histogram of the differences in days (dt) versus the differences in magnitude (dm). They proposed an image-based classification of variable stars data taken from the Catalina Real-Time Transient Survey. They obtain performances comparable to Random Forest (RF;\cite{2001MachL..45....5B}) without feature extraction, and highlight potential future applications with this approach.\newline
More recently, \cite{2020ApJ...897L..12S, 2022ApJ...938...37S} studied the classification of phasefolded LCs in OGLE data. They represented the LCs as 8-bit images with a size of 128 × 128 pixels, using a black background with white plotted dots. The main idea is to simulate the traditional approach to displaying and analyzing time-series data. In their first paper, they presented the methodology for image-based classification, and in their second paper, they extended their work by using a multi-input neural network that combined images with tabular information.\newline

This work is the first in a series focusing on the highly accurate classification of variable stars in big data astronomy. The aim of this initial paper is to further explore image classification. We present the methodology and demonstrate that this approach is suitable in terms of speed, accuracy, and minimizing computing resources. The second paper will concentrate on the adaptability of the model to various surveys, emphasizing the optimization of the number of classified stars necessary for retraining the model in each survey. \newline
The paper is organized as follows:  in Sect. \ref{sec:data} we summarize the OGLE data from different survey missions and explain the download process. We also describe the preprocessing steps used to generate training, validation, and test sets. In Sect. \ref{sec:Methodology} we present our Convolutional Neural Network architecture and detail the training process. In Sect. \ref{sec:result} we present the main results for the classification of variable stars with different balancing techniques. We also propose a combination of tabular and image information to improve our convolutional neural network. In Sect. \ref{sec:discussion} we compare our algorithm with others and discuss the computational resources and time required. Finally, in Sect. \ref{sec:summary}, we summarize our work and present future projections on this topic.

\begin{table*}
\caption{Number of variable stars used in this study from OGLE III and IV time series.}
\label{table:1}
\centering
\begin{tabular}{c c c c c c c c} 
 \hline \hline
 \makecell{Variability class} & \makecell{Acronym} & \makecell{Non-unique LC from \\ OGLE Catalogs} & \makecell{I Filter \\ Data Missing} & \makecell{$n_{obs}<60$} & \makecell{Selected from \\ OGLE IV} & \makecell{Possible \\ blended star} & \makecell{Final \\ numbers} \\ 
\hline
Ellipsoidal binary & ELL & 26880 & 925 & 29 & 0 & 510 & 25416 \\
Mira & M & 74542 & 1214 & 4851 & 5161 & 36 & 63280 \\
Classical Cepheids & CEP & 19680 & 116 & 481 & 7638 & 79 & 11366 \\
Delta Scuti & DST & 45413 & 532 & 483 & 47 & 4344 & 40007 \\
Eclipsing binaries & E & 516143 & 14273 & 699 & 29381 & 866 & 470924 \\
Long period variable & LPV & 335255 & 0 & 647 & 122 & 1055 & 333431 \\
RR Lyrae & RR & 170408 & 1466 & 3332 & 41421 & 300 & 123889 \\ 
 \hline
 Total & - & 1188321 & 18526 & 10522 & 83770 & 7190 & 1068313 \\ 
 \hline
\end{tabular}
\tablefoot{The Col. 1 presents the class of variability. The Col. 3 displays the number of variable stars obtained from the OGLE catalogs. The Col. 4 lists stars without time series in the I-band filter. The Col. 5 indicates stars with fewer than 60 observations.The Col. 6 shows stars that are observed in both OGLE III and OGLE IV. The Col. 7 shows stars eliminated due to an internal match within 1 arcsec. The Col. 8 presents the final count of independent stars. Table \ref{table:references_for_classes} shows the OGLE reference for each variability class.}
\end{table*}

\section{Data}\label{sec:data}
\subsection{Data and pre-procesing}
For the development of this work, we utilized the time series data from OGLE. The main objective of OGLE is the search for dark matter with microlensing phenomena \citep{1992AcA....42..253U}. However, due to the extensive number of observations, and the prolonged timeframe that can be archived over half a century, there has also been a focus on the identification and classification of variable stars. OGLE has monitored the Large Magellanic Cloud, the Small Magellanic Cloud, the Milky Way disk, and the Milky Way bulge. \newline
The observations were conducted with the 1.3-m Warsaw telescope at Las Campanas Observatory in Chile. The photometry was obtained in the I band, which is close to the standard Kron-Cousins system, and in the V band, similar to the Johnson V photometric band \citep{2015AcA....65....1U}\footnote{\href{http://svo2.cab.inta-csic.es/theory/fps/index.php?mode=browse&gname=LCO&gname2=OGLE-IV}{SVO FPS Carlos Rodrigo}}. The majority of the observations were conducted in the I band filter, the specific proportion varying depending on the field. For instance, in OGLE III for the Small Magellanic Cloud, $90\%$ of the observations were in the I band \citep{2013AcA....63..323P}. The total number of epochs also varies depending on the field. For instance, in the I band, OGLE III exhibited a range from several dozen to approximately 3000 measurements \citep{2011AcA....61..285S}. In OGLE IV, the number ranged from 100 to over 750 in the I band and from several to 260 in the V band \citep{2015AcA....65..297S}. From the information obtained for this work, we reported a cadence ranging from $\sim$0.05 days to $\sim$216 days, and a baseline ranging from $\sim$30 days to $\sim$4500 days. \newline
From 2001 to 2009,
OGLE III observations were carried on with an eight-CCD detector mosaic camera featuring a pixel scale of 0.26 arcsec/pixel and a field of view spanning 35x35 arcminutes. During 2010-2015, OGLE IV employed a 32-chip mosaic camera, which covered approximately 1.4 square degrees of the sky. In its final stage, OGLE achieved coverage of 3000 square degrees in the Galactic disk and bulge \citep{2020AcA....70..101S}, along with 650 square degrees in the Magellanic Clouds \citep{2016AcA....66..421P}, observing 70 million stars in the I band \citep{2023AcA....73..105S} with a magnitude range of 10-21.7 \citep{2015AcA....65....1U}. \newline
The OGLE photometric data products are obtained according to the Image Subtraction Analysis method \citep{1998ApJ...503..325A} and was implemented by \cite{2000AcA....50..421W}. The OGLE team has conducted a massive search for periodic signals in time series data. In most of their work, they used the code \texttt{FNPEAKS}, written by Z. Kołaczkowski\footnote{\href{http://helas.astro.uni.wroc.pl/deliverables.php?lang=en&active=fnpeaks}{FNPEAKS}}. This algorithm employs a Discrete Fourier Transform to identify the most significant periods, incorporating amplitude and signal-to-noise ratio information. In some studies \citep{2013AcA....63..323P, 2016AcA....66..405S, 2022ApJS..260...46I}, OGLE team combines this algorithm with others, such as the \texttt{Analysis of Variance-based method} \citep{1996ApJ...460L.107S}, the \texttt{Box-Least Squares algorithm} \citep{2002A&A...391..369K}, and the \texttt{Lomb–Scargle periodogram} (\citet{1976Ap&SS..39..447L}, \citet{1982ApJ...263..835S}). In \cite{2011AcA....61..103G}, instead of using \texttt{FNPEAKS}, the authors employed the \texttt{phase dispersion minimization} \citep{1978ApJ...224..953S} and the \texttt{string-length method} \citep{1965ApJS...11..216L}.

To obtain the LCs, we started by downloading\footnote{\href{https://www.astrouw.edu.pl/ogle/}{https://www.astrouw.edu.pl/ogle/}} all catalogs of variables stars found in OGLE III and IV. Data were divided by mission (I, II, III and IV), field, and variability class assigned to each object in the respective field. We selected missions III and IV because of the large number of variable stars classified and the homogeneity of the classes. We downloaded three files, one for each variability class. Those files were called ident.dat, variability\_class.dat (e.g., ecl.dat for eclipsing binary), and phot.tar.gz. From ident.dat, we obtained the identifier (ID) designated by OGLE, the right ascension, and the declination. This ID is unique for each star, although it was possible to observe the same stars in different OGLE missions, resulting in two time series for the same star. From variability\_class.dat we obtained the ID and the periods. From phot.tar.gz we obtained the time series of the stars. This compressed archive contained two folders, I and V with the time series in the respective filter. We only used observations in the I band, due to its larger number of observations. We used IDs to cross-match the coordinates with the periods and time series of the stars.\newline
We selected variability classes with the highest number of classified stars to ensure the data set was as balanced as possible and to include as many examples as possible. The selected classes include Classical Cepheid (CEP), RR Lyrae (RR), Long Period Variable (LPV), Miras (M), Ellipsoidal Binary (ELL), Delta Scuti (DST), and eclipsing binary (E) (for acronyms, see Column 2 of Table \ref{table:1}). As in OGLE IV, we also considered M and ELL as main variability classes, and extended this criterion also to the OGLE III database, in which they were originally classified as a subgroup of LPV and E, respectively.\newline
We noted that the ``ident.dat'' file lists $18526$ stars that did not have associated LCs in the ``phot.tar.gz'' file. We also attempted to download these time series directly from the online folder, but found that these stars were absent there as well. Table \ref{table:1}, col. 3, shows the final number of stars we successfully downloaded.\newline
We filtered the catalog to ensure that only one time series per star was included. In instances where we had access to both OGLE III and OGLE IV LCs, we exclusively utilized those from OGLE IV. Thus guaranteeing the independence of the training, validation, and test sets (see Sect. \ref{sec:Splitdata}). We decided not to include OGLE III LCs as independent time series because our objective was to focus on unique stars. Additionally, we did not combine time series from OGLE III and OGLE IV, as doing so could potentially introduce biases into the subset of stars with a larger observation baseline. \newline
We aimed to ensure with high certainty that our sample contains no blended stars. In order to discard LCs coming from potentially blended sources, we adopted a very conservative threshold of three elements of resolution. Meaning that we discarded sources with an additional ``companion'' within $1.6$ arcseconds. The pixel scale of OGLE camera is $0.26$ arcseconds per pixel. The median seeing, approximated by the Full Width at Half Maximum (FWHM) of the stellar Point Spread Function (PSF) measured in dense stellar fields, was about $1.25$ arcseconds \citep{2015AcA....65....1U}. When using differential imagining techniques, one could obtain clean LCs for objects separated less than three times the spatial resolution. However, this assumption relied on only one of the two sources not being variable above the noise limit. Since we could not conduct dedicated inspection of close by projected companions, we adopted the previously mentioned ``three elements of resolution'' to guarantee that in the rest of the study we would be dealing with ``non-contaminated'' LCs. To identify stars that were closer than 1.6 arcseconds to each other, we employed the internal match feature of \texttt{Topcat} \citep{2005ASPC..347...29T} to search within the sample coordinates. Our goal is to identify typical variability phenomenon for each class, leading us to filter out atypical cases. The stars that were removed are listed in Table \ref{table:1}, column 6. \newline
We applied sigma-clipping with a factor 3. On average, this procedure ejected approximately $0.65\%$ of the observations from the entire time series, predominantly removing outliers. After visual inspection of a subsample of stars, the concern was raised on the impact of in homogeneity of number of observations and its impact in classification. To mitigate this possible bias, we analyzed the distribution of number of observations of our sample. ``Under sampled'' LCs, meaning those with less than 60 observations, only represented $1\%$ of the first quintile were discarded, and ``oversampled'' LCs, those with more than 2000 observations (the $11\%$ most sampled objects), were downsampled randomly to achieve a more uniform density distribution of observation. The typical baseline of objects with a higher density of observations was similar to that of objects with fewer observations. Subsampling objects with a greater number of observations probed the same timescales as objects in both domains. Therefore, before resampling, the minimum and maximum cadences were $\sim$0.16 days and $\sim$2.25 days, respectively. After resampling, the cadence was $\sim$0.24 days to $\sim$2.25 days.
\section{Methodology}\label{sec:Methodology}

\subsection{Data representation}
In the phot.tar.gz files downloaded from OGLE, an original time series consisted of Heliocentric Julian Day - 2450000, the magnitude of the star, and the uncertainty of the magnitude. We used the variability period calculated by OGLE team to display the LC in phase. In this way, we visualize the variability of the star in one cycle of its period with Eq. \ref{eqn:ecuacion1}.
\begin{equation}
    \phi = \frac{t-t'}{P} - int[\frac{(t-t')}{P}]
     \label{eqn:ecuacion1}
\end{equation}
\noindent where $\phi$ is the phase, $t$ is the time at which the measurement was made, $P$ is the period and $t'$ is an arbitrary epoch. \newline
\begin{figure}
  \begin{center}
    \includegraphics[width=2.9in]{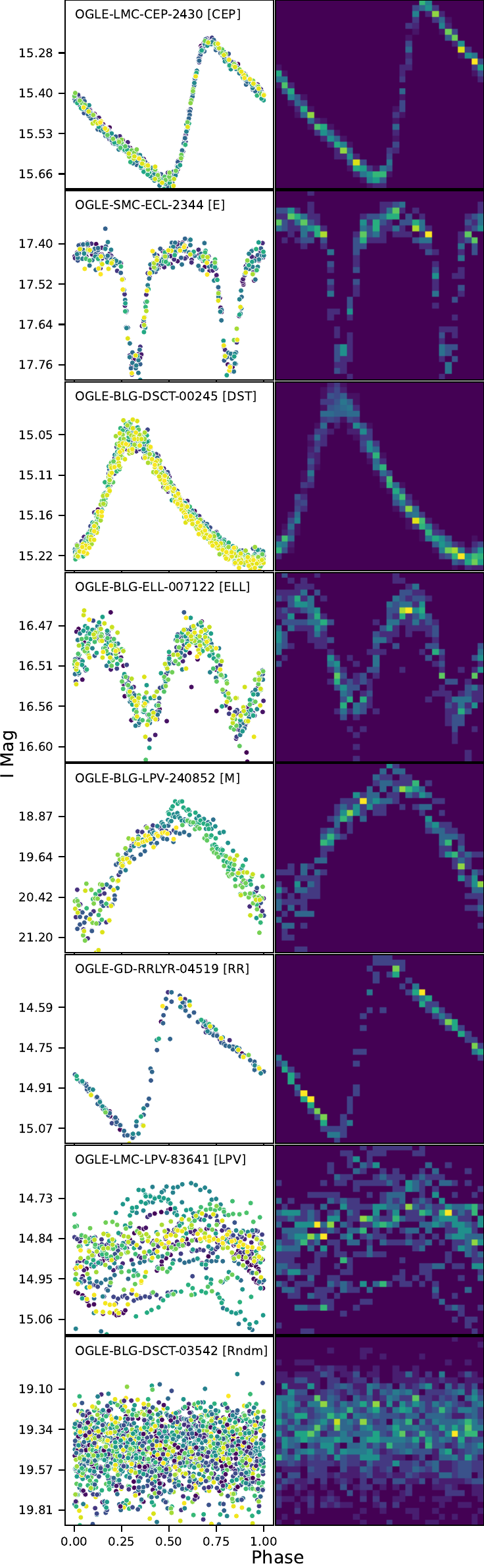}
  \end{center}

  \caption{\small Phase LCs for the
different variability classes considered in the
OGLE sample. For each class, the first column shows the lightcurve with colors showing the different cycles. From the top the different variability classes are M, CEP, ELL, E, DST, RR, LPV, and the Random period Class. The second columns are the 32x32 histogram and the color represent the number of observation in each bin with a min-max normalization.}
  \label{fig:fase_hist}
\end{figure}

\begin{table*}
\begin{center}
      \caption{\small Period and amplitude for different variability classes.}
  \label{table:amplitude_period}
\begin{tabular}{ccccccc}
 \hline
 &  & Number of Objects & D$_{\rm 1 Period}$ [d] &  D$_{\rm 9 Period}$ [d] &  D$_{\rm 1 Amplitude}$ [$\Delta$ I$_{\rm Imag}$] &  D$_{9 Amplitude}$  [$\Delta$ I$_{\rm Imag}$] \\
\cline{1-7}
Variability class & Field &  &  &  &  &  \\
\cline{1-7}
\midrule
\multirow[t]{3}{*}{ELL} & BLG & 10723 & 0.46 & 151.1 & 0.07 & 0.28 \\
 & LMC & 515 & 0.8 & 84.02 & 0.09 & 0.37 \\
 & SMC & 128 & 0.74 & 10.11 & 0.12 & 0.3 \\
\cline{1-7}
\multirow[t]{4}{*}{M} & BLG & 6714 & 192.73 & 453.94 & 1.56 & 4.0 \\
 & GD & 4280 & 226.46 & 490.01 & 1.65 & 4.05 \\
 & LMC & 302 & 205.07 & 552.18 & 1.48 & 4.11 \\
 & SMC & 70 & 265.73 & 546.74 & 1.76 & 4.34 \\
\cline{1-7}
\multirow[t]{4}{*}{CEP} & BLG & 180 & 0.37 & 14.27 & 0.2 & 0.83 \\
 & GD & 1631 & 0.94 & 12.86 & 0.21 & 0.7 \\
 & LMC & 4606 & 0.9 & 5.58 & 0.2 & 0.57 \\
 & SMC & 4949 & 0.78 & 4.21 & 0.24 & 0.75 \\
\cline{1-7}
\multirow[t]{4}{*}{DST} & BLG & 4621 & 0.05 & 0.12 & 0.16 & 0.59 \\
 & GD & 2120 & 0.06 & 0.16 & 0.1 & 0.49 \\
 & LMC & 3859 & 0.06 & 0.1 & 0.48 & 1.34 \\
 & SMC & 766 & 0.06 & 0.1 & 0.64 & 1.59 \\
\cline{1-7}
\multirow[t]{4}{*}{E} & BLG & 9895 & 0.34 & 4.3 & 0.24 & 1.03 \\
 & GD & 300 & 0.28 & 2.77 & 0.16 & 1.28 \\
 & LMC & 964 & 1.09 & 17.09 & 0.15 & 0.95 \\
 & SMC & 207 & 0.78 & 26.87 & 0.14 & 0.94 \\
\cline{1-7}
\multirow[t]{3}{*}{LPV} & BLG & 7621 & 11.16 & 92.16 & 0.05 & 0.46 \\
 & LMC & 3073 & 13.64 & 339.82 & 0.04 & 0.4 \\
 & SMC & 672 & 14.26 & 382.68 & 0.05 & 0.35 \\
\cline{1-7}
\multirow[t]{4}{*}{RR} & BLG & 6171 & 0.29 & 0.64 & 0.29 & 0.9 \\
 & GD & 918 & 0.3 & 0.65 & 0.31 & 0.94 \\
 & LMC & 3669 & 0.32 & 0.65 & 0.43 & 0.96 \\
 & SMC & 608 & 0.37 & 0.66 & 0.51 & 0.96 \\
\cline{1-7}
\bottomrule
\end{tabular}
\end{center}
\tablefoot{ Values calculated using the final preprocessed sample. We present the D1, D9, and median values for both amplitude and period. D1 corresponds to the first decile, and D9 to the ninth decile of the sample.}
\end{table*}
We phasefolded the LCs of the objects belonging to
the seven variability classes described previously. In addition to the nominal classes in the OGLE taxonomy, we artificially populated a ``spurious class'', by randomly resampling real LCs to arbitrary periods drawn from each class period distribution. We included this class to address a potential weakness affecting phasefolded LCs: the assumption of accurate and correct periods. If the period was inaccurate, the phasefolded image had significant scatter and showed a smooth behavior through phase. Including spurious class examples in the training set enabled the model to identify inaccurate period calculations, and, simultaneously, prevented classifying them wrongly into one of the other classes. Table \ref{table:amplitude_period} presented the period and amplitude domains for each variability class. We presented the first decile, $D_{1}$, and the ninth decile, $D_{9}$, corresponding to the lower 10\% and upper 10\% of the sample, respectively. We also categorized by different environments: Galactic (Bulge and Disk) and the Magellanic Clouds (Large and Small). We were aware of variations in LCs due to environmental factors such as metallicity. Nonetheless, we aggregated different fields to create a generalized sample for our model.

We used the phase and the magnitude of the stars to represent the LC using a two-dimensional histogram of size 32x32. In this histogram, 32 bins ranging from 0 to 1 are allocated for the phase, and another 32 bins cover the range from the minimum to the maximum magnitude of each star. We then normalized the histogram to scale the bin counts to values between 0 and 1 by dividing by the histogram's maximum count. Although increasing the histogram size to 64x64 could potentially have improved classification, we chose 32x32 images to analyze more LCs with less memory, as doubling the resolution would have quadrupled the memory usage.

Figure \ref{fig:fase_hist} shows an example of the eight different classes of variable star selected from OGLE data.\newline
\newline
\begin{figure*}
  \begin{center}
    \includegraphics[scale=0.75]{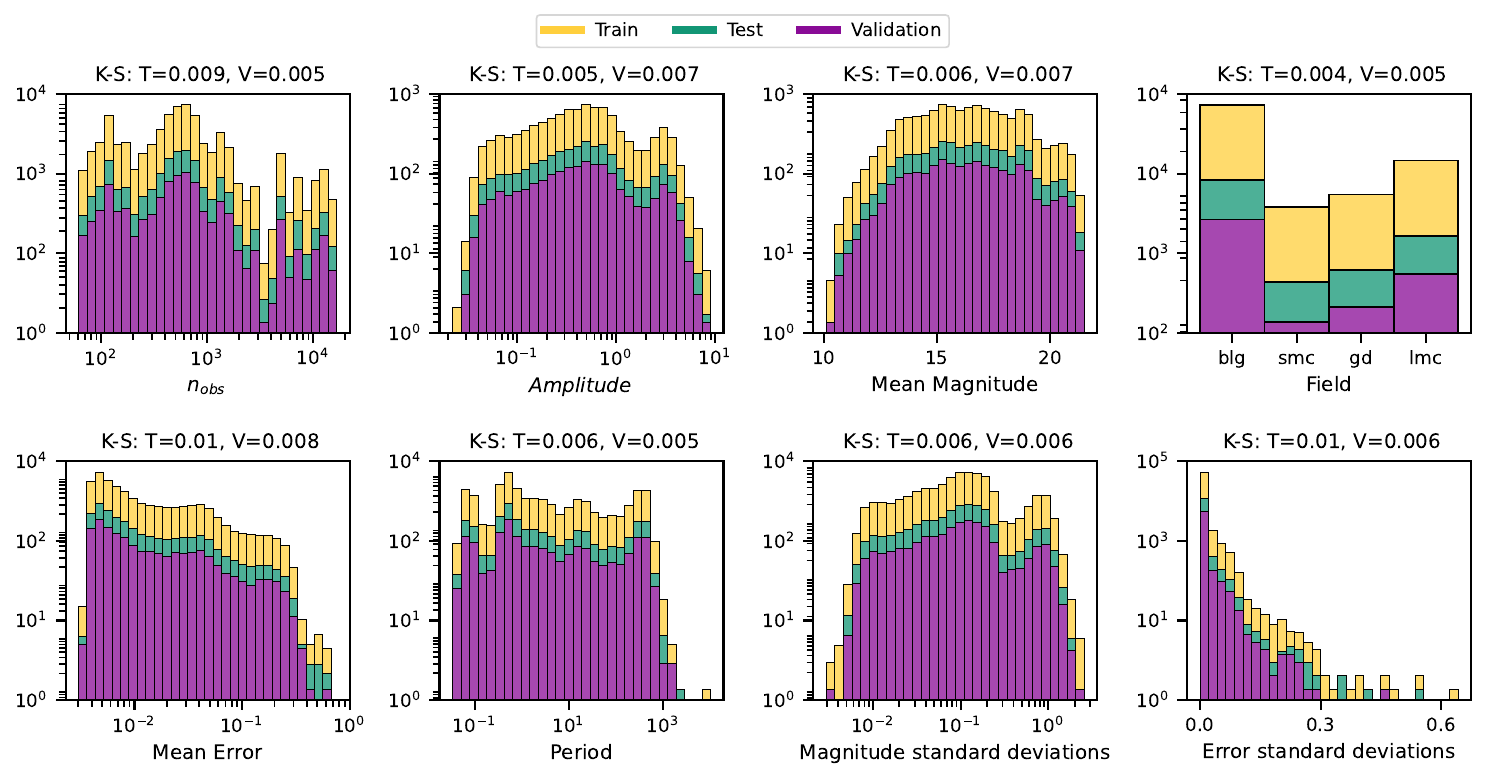}
  \end{center}

  \caption{\small Distribution of observational parameters of the variable stars used in this work is shown. The number of observations, amplitude, mean magnitude, field, mean error, period, magnitude standard deviation, and the standard deviation of magnitude error are presented. The yellow, purple, and green colors represent the training, validation, and testing sets, respectively. The distributions are stacked for better visualization, and the titles correspond to the KS test for validation and testing.}
  \label{fig:parameter_distribution}
\end{figure*}
\subsection{Train validation and test sets}\label{sec:Splitdata}
We separated the OGLE data to train the model into three sets: Training, Validation, and Test sets in a proportion of 70:15:15. The training set was used by the algorithm to learn directly how to recognize different labeled data. The validation set was utilized indirectly to optimize the learning algorithm, aiming to quantify its level of generalization. Finally, the Test set, which was entirely independent, served to evaluate the algorithm's classification performance, as these data had not been previously exposed to the model, either directly or indirectly.\newline
First, we created a balanced dataset by reducing the sample size to 11 366 stars per class, matching the population of the CEP, which was the least represented. This was achieved through Undersampling (U) (see \ref{sec:balanceddata}). We had 11 366 independent CEP, which were divided into 8 404, 1484, and 1478 to train, validate, and test, respectively. We selected the same number of examples for the rest of the classes. The total number of stars for training, validation, and testing were 67 232, 11 872, and 11 824, respectively, across eight classes: CEP, RR, L, M, ELL, DST, E, and the spurious class. For the spurious class, we selected subsets of time series from each class and chose a randomly resampled period. It is important to note that these selected time series originated from real observations, not augmented stars. Additionally, we ensured that time series from different sets were not combined. As the training set had been balanced through U, resulting in approximately 8 000 stars per class, it was hereafter referred to as 'Train-8'.

We evaluated the representativeness of the balanced datasets by comparing the distribution of the observables and physical parameters for the samples. These comparisons are useful for characterizing the sample as astronomical objects and for assessing the representativeness of the entire algorithm's flow. In Fig. \ref{fig:parameter_distribution}, we present the distributions of various parameters across different sets. The parameters include the number of observations, amplitude, mean magnitude, field, mean error, period, magnitude standard deviations, and error magnitude standard deviations for each star. The title of each distribution show the Kolmogorov–Smirnov (K–S) test \citep{kolmogorov1933sulla, smirnov1948table} for training set with validation and test. Owing to the combination of different classes and environments, the properties of each star exhibit diversity. Nevertheless, it is observable from the Fig. \ref{fig:parameter_distribution} that similar distributions prevail in the three sets for most of the distribution range. Minor discrepancies are noted only in the extremes of the magnitude standard deviations and error standard deviations.
\begin{figure}
  \begin{center}
    \includegraphics[scale=0.35]{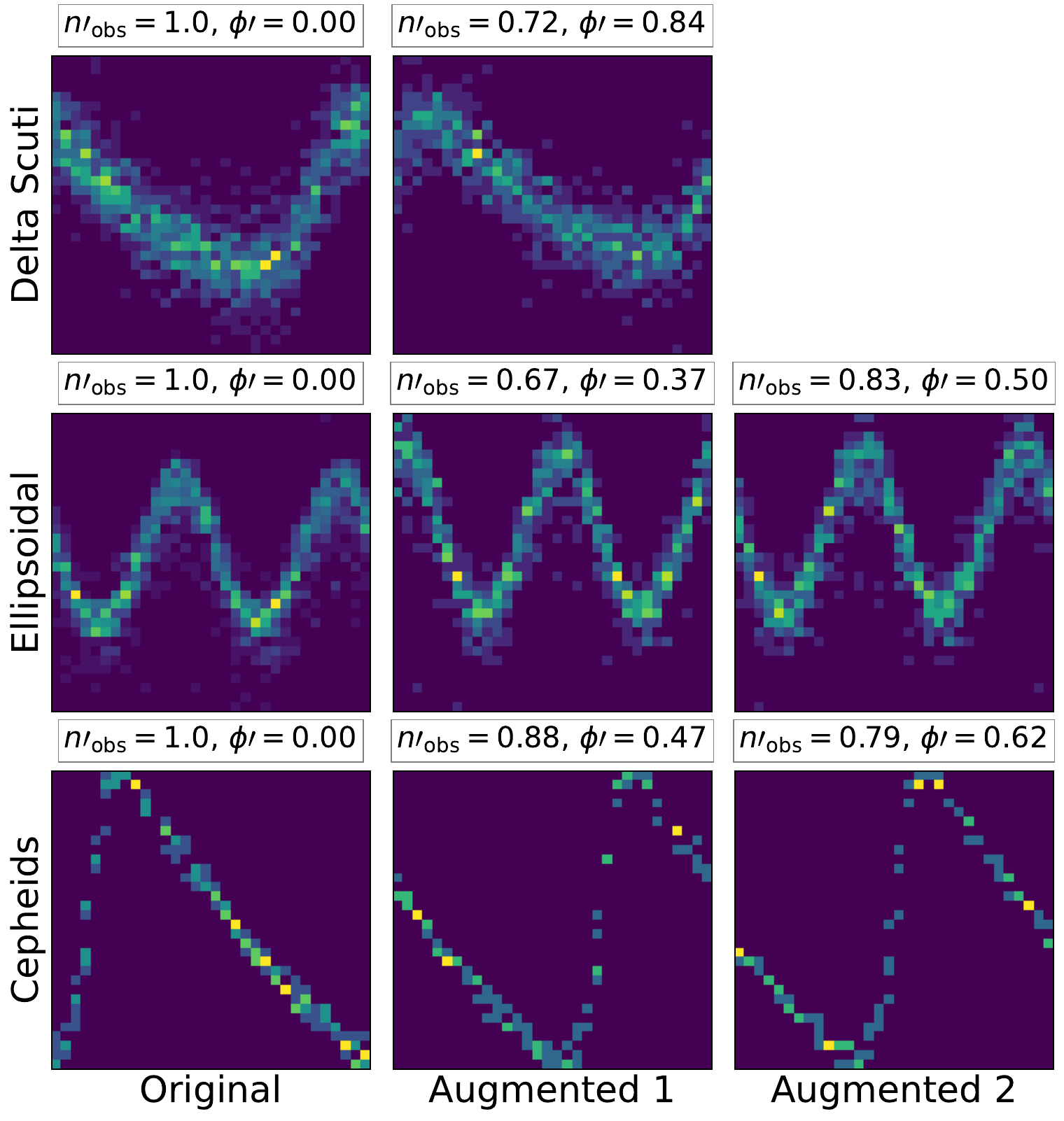}
  \end{center}

  \caption{\small Examples of DA for CEP, DST, and ELL are shown. The first column represents the original LC, while the other columns represent different modifications of the LC. At the top of each image, we display the fraction of the total number of observation ($n'_{obs}$), and the offset in phase ($\phi'$).}
  \label{fig:data_augmented}
\end{figure}

\begin{figure*}
  \begin{center}
    \includegraphics[width=6in]{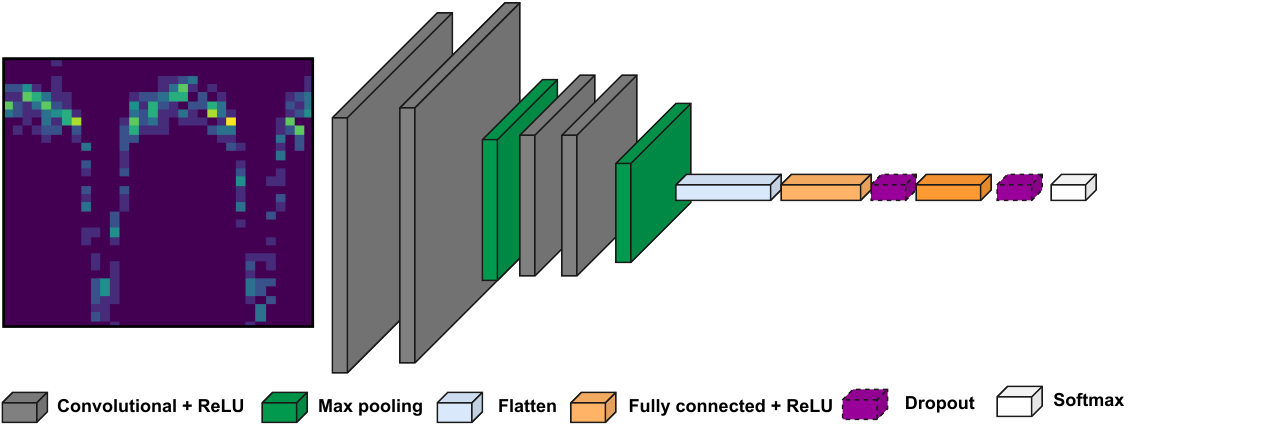}
  \end{center}

  \caption{\small Representation of our convolutional neural network architecture. We show a 2D histogram of an E as an input example. Different layers of the network are displayed in various colors. The network is composed of 2 blocks of convolution layers, convolution layer, max pooling, and dropout. Then, we flatten the input and use two fully connected layers followed by the output layer. The output consists of a neuron with softmax activation. Table \ref{table:hyper_parameters} show the hyperparameters used in this architecture.}
  \label{fig:architecture}
\end{figure*}
\subsubsection{Balanced Data}\label{sec:balanceddata}
In astronomy, we often encounter imbalanced datasets characterized by widely differing numbers of classified objects for several variability classes. The distribution of astrophysical phenomena in the universe is not uniform. This non-uniformity arises either from inherent biases towards intrinsically rare phenomena or from observational factors that skew detections towards specific phenomena. In the classification of variable stars, this issue becomes evident (see Table \ref{table:1}). E (the majority class) outnumber CEP (the minority class) by a factor of 40. This imbalance can skew standard classifiers to be overwhelmed by the larger classes and to ignore the minority class \citep{10.1145/1007730.1007733}. To deal with this problem, we employ different approaches to balance the data: Undersampling (U), Batch Balancing (BB), and Data Augmentation (DA). \newline
The U technique represents a popular approach for addressing the class imbalance problem. This technique involves training only with a subset of the objects populating the majority classes. The size of this subset is chosen according to the availability for training and testing of the minority classes. This approach makes it a straightforward and efficient method for handling imbalanced datasets. The main drawback is that it ignore many examples of the majority class \citep{4717268}.
However, if we have a representative subsample of examples, we can expect favorable outcomes from this approach.\newline
The BB involves dividing the data used to train the model into subsamples, known as batches. These batches are randomly selected without repetition from the training set \ref{sec:Splitdata}. This procedure continues until all the data from the training set has been processed. One epoch is completed when the entire training set has been used. After completing one epoch, the entire training set is reintroduced, and the process begins again. In the BB training procedure, as described in \cite{8665709}, the minority class is repeated within an epoch to ensure balanced batches. \newline 

Finally, the DA consists of generating synthetic data derived from the original data set. It is imperative that this technique is applied solely to the training set. The objective is to create synthetic data that, although different from the original, preserves relevant patterns from the data. Typical image data augmentation techniques are rotations, flips, cropping, and blurring, but they are not suitable in our context as they could inadvertently alter the intrinsic astrophysical behavior of stars \citep{2020ApJ...897L..12S}. An alternative is to utilize a generative model to produce synthetic LCs, as demonstrated by \citep{2022AJ....164..263M} using OGLE III LCs and \textit{Gaia} DR2 stellar parameters. However, we opted not to use this model due to potential data leakage risks. Nonetheless, their approach to DA is similar to ours. We implemented DA with three different variations in the LC: magnitude shift, phase shift, and reduction in the number of observations variations. \newline
The magnitude shift was introduced by \citep{2020ApJ...897L..12S} who presented this methodology as a DA. This approach does not account for any correlated noise and assumes the observations are independent and identically distributed random variables, with is not necessarily the case. We complement the modifications of the LC with the other three changes. The magnitude shift consists of the generation of Gaussian noise with a mean of zero and a standard deviation equal to the observational error; then we aggregate this error to the magnitude. For phase variations, we change the value $t'$ in equation \ref{eqn:ecuacion1}. We maximize the spaced numbers of $t'$ over the interval $ 0+1/32<t'<1-1/32$ according to the augmented LC. We created a binned LC for the reduction in the number of observations. The binning method consists of grouping data in continuous intervals and calculating a mean phase and mean magnitude per interval. We randomly chose the bin values between $0.5n_{obs}$ and $0.9n_{obs}$. For stars for which the resampling results in fewer than 60 observations, we randomly chose between 60 and the number of observations of the star. \newline
We present an example of DA in Fig. \ref{fig:data_augmented}. This shows DA for the three variability classes that require augmentation to achieve balanced data.

\subsubsection{Balanced Training Sets}

\begin{table*}[htpb]
  \begin{center}
    \caption{Summary of different balanced training sets.}
    \label{table:Trainings_sets}
    \begin{tabular}{ccccc} 
   
   \hline

Training Set  & Star per Class & balanced technique & Minority class & Total Number \\
&  &  \\

\hline

\hline
 
   Train-8 U     & 8404 & Undersampling & CEP & 90928  \\
   Train-22 DA    & 22454 & Data Augmentation   & ELL & 179632 \\
   Train-22 BB    & - & Batch Balanced   & ELL & 165582  \\
   Train-37 DA & 37045 & Data Augmentation  & DST & 296360  \\
   Train-37 BB & - & Batch Balanced  & DST & 253128 \\
   Train-60 DA     & 60318 & Data Augmentation & M & 482544 \\
   Train-60 BB     & - & Batch Balanced & M & 369493 \\
\hline
  \end{tabular}
  \end{center}
\end{table*}

We created seven different training sets to investigate the limitations of the balancing techniques. Using data from Train-8 and previously unused data, we added new data to Train-8 when available. We decided the number of stars in each class based on the next minority class, aiming for using all data. For the different sets we employed BB and DA. The Table \ref{table:Trainings_sets} presents a summary of the training set. The number in the name of the training set corresponds to the number of stars per class. The absence of information in the column stars per class is due to the lack of a fixed number of stars per class. \newline

\subsection{Convolutional neural network}
Convolutional Neural Networks (CNNs) are a type of Artificial Neural Network (NN) designed to process data in the form of fixed-size multiple arrays, for example: 1d sequence, 2D images, 3d videos, among others \citep{2015Natur.521..436L}. The basic architecture of CNNs consists of four building blocks: convolution layers, regularization layers, pooling, and fully connected layers. The first three layers are used to extract the relevant features from the data, and the last layer learns the complex relationships between the features and the classes / labels. In this work, we utilized an architecture similar to that described by \cite{2020ApJ...897L..12S}, optimized for our input data and with fewer parameters. In this section, we provide a qualitative description of our architecture (see Fig. \ref{fig:architecture}). For a comprehensive review of CNN fundamentals, we recommend \cite{cong2023review}.\newline
The convolution layer is the crucial component of the CNN. These layers are composed of free-learned parameters named kernels or filters of a given size. These parameters perform the convolution operation, which is a dot product between the input array and the filter, producing a feature map that summarizes the input array. During the convolution operation, we move the filter to apply the convolution across the entire array. The Padding and strides are parameters in convolution layers. Padding involves adding zero values to the array borders, while ``same'' padding adds the necessary zero values to ensure the convolved array retains the shape of the input array. This technique prevents the loss of border information. The stride parameter determines how many steps we move the filter. \newline
The pooling, regularization, and fully connected layers are useful for analyzing input arrays. The pooling layer aims to decrease the input size without losing essential information. Max pooling involves selecting a region in the image and keeping only the highest value. The regularization layers is used to increase the generalization of the network. The dropout is a type of regularization layers, in this technique, neurons are randomly selected and ignored during the training process \citep{JMLR:v15:srivastava14a}. The fully connected layer is a regular NN. This is a block compound with units or neurons; each unit works in parallel and is connected to all neurons in the previous layer. Each neuron received the output of all previous neurons and used this information to calculate an activated value. The activation functions are a fundamental part of the NN. It is utilized to aggregate nonlinearity to the output that enables the network to solve complex nonlinear problems. The idea is to apply some mathematical nonlinear function $(\delta)$ to the input $(x)$ and get the nonlinear output $\delta(x)$. Many different activation function exists, and their choice translate in changes in the network's performance \citep{app10051897}. \newline
In our model, the input to the CNN was the 2D histogram of the phasefolded LC. Since we used a single channel (single observational band), the dimensions of the input layer were equal to the input array of 32x32x1. We incorporated two convolutional layers with 32 filters, the Rectified Linear Unit (ReLU) activation function, and ``same'' padding. We used a stride of one, meaning the filter moved one position at a time. After the two convolution layers, we applied max-pooling with a 2x2 kernel. We repeated two blocks of convolution, convolution, and Max pooling, then we aggregated two fully connected layers, regularization layers, and the output layer. We defined 2 layers with 1024 and 512 neurons. Between the fully connected layers, we inserted a dropout as regularization layers. Finally, we defined eight neurons with the softmax activation function for the output layer. The softmax function, with one neuron per class, facilitated multiclass classification. It outputted a value between 0 and 1, which represented the classification probability for each class. \newline
This study employs a relatively standard CNN architecture, hence, its optimization is out of scope. Within this architecture, we employed the most commonly used activation functions: ReLU in the hidden layers and softmax in the output layer.
\begin{figure*}
  \begin{center}
    \includegraphics[width=7in]{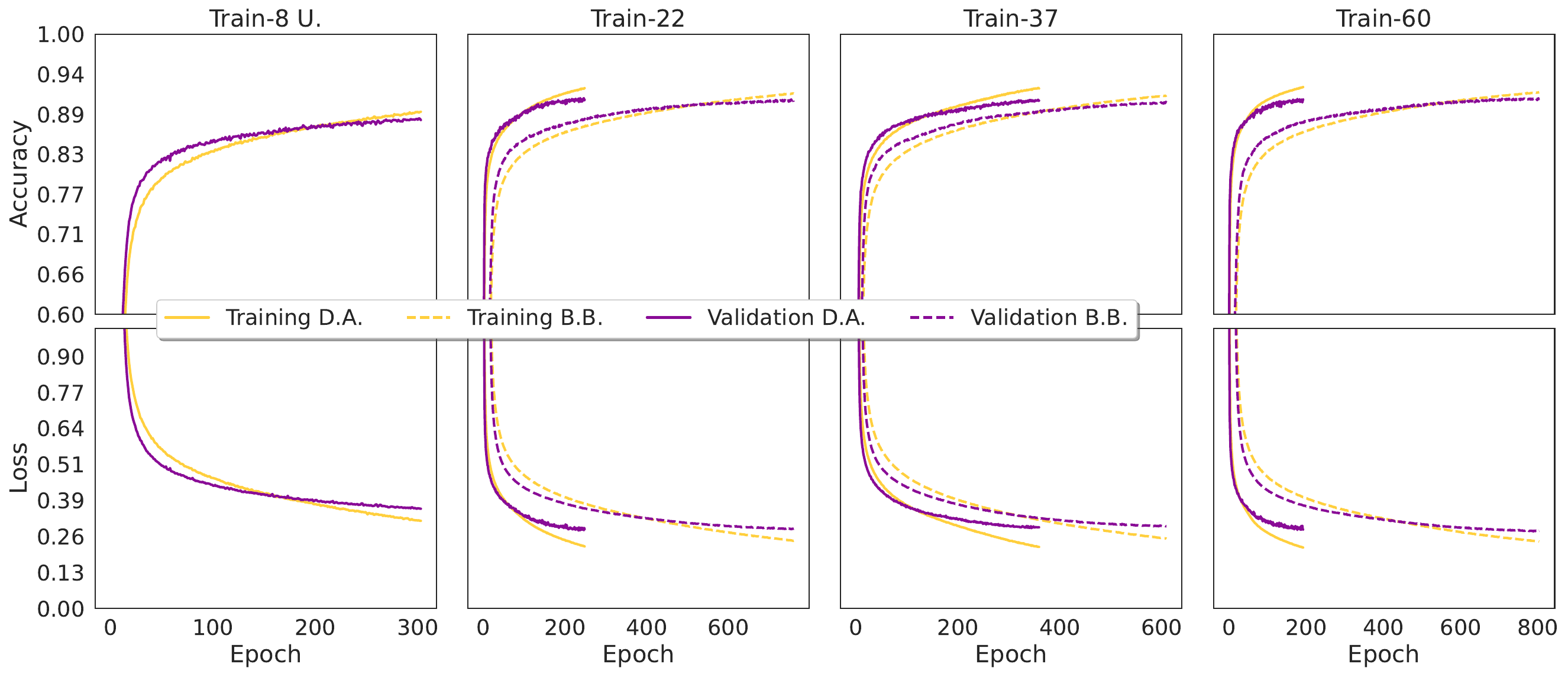}
  \end{center}
  \caption{\small Training process for the convolutional neural networks. The first and second rows show the accuracy and loss, respectively. The behavior of the training set is shown in yellow, while the behavior of the validation set is shown in purple. The training process is stopped at 1000 epochs or when the network starts performing better on the training set than on the validation set.}
  \label{fig:training}
\end{figure*}
 \newline
\begin{figure}
  \begin{center}
    \includegraphics[scale=0.51]{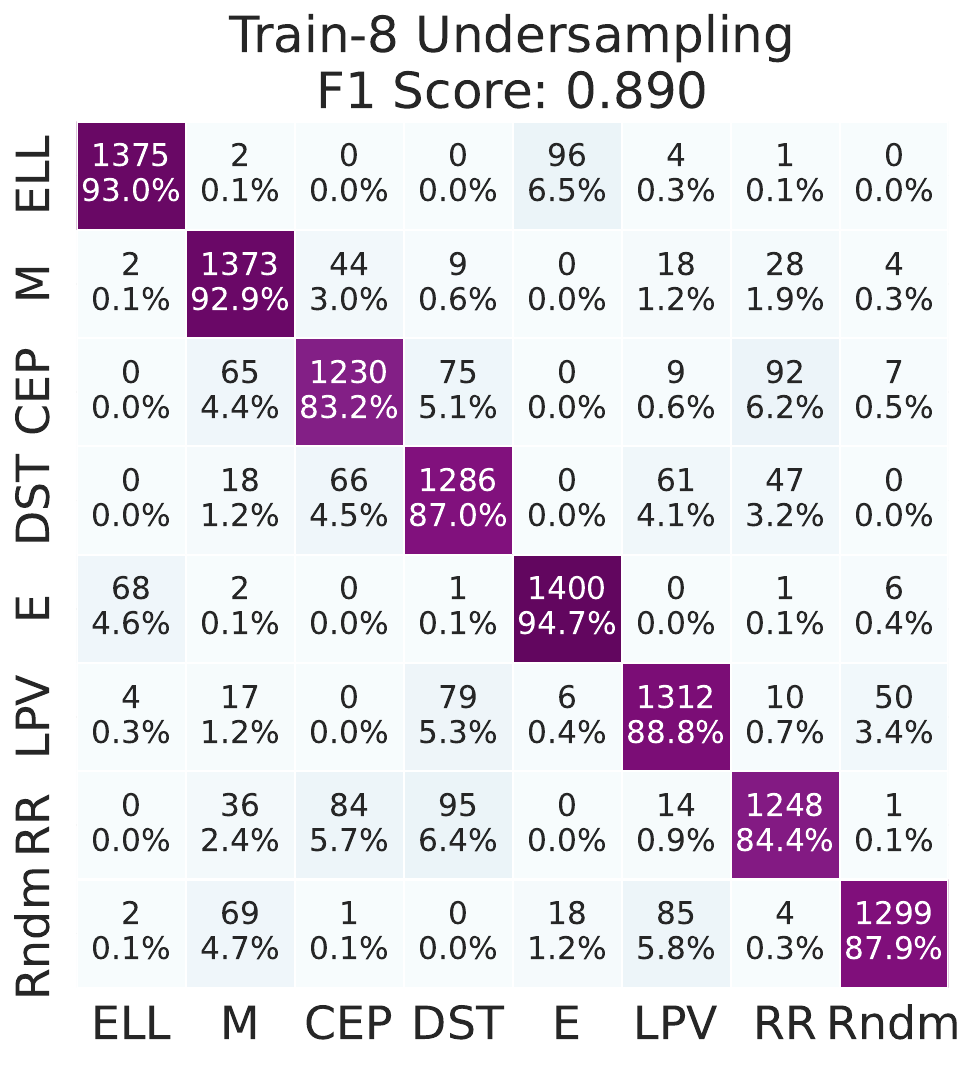}
  \end{center}

  \caption{\small Confusion matrix for the Train-8 test dataset, balanced via U, illustrating initial CNN performance.}
  \label{fig:CNN_Undersampling}
\end{figure}
 \subsection{Training process}\label{subsec:training}
 \begin{figure*}
  \begin{center}
    \includegraphics[width=7.1in]{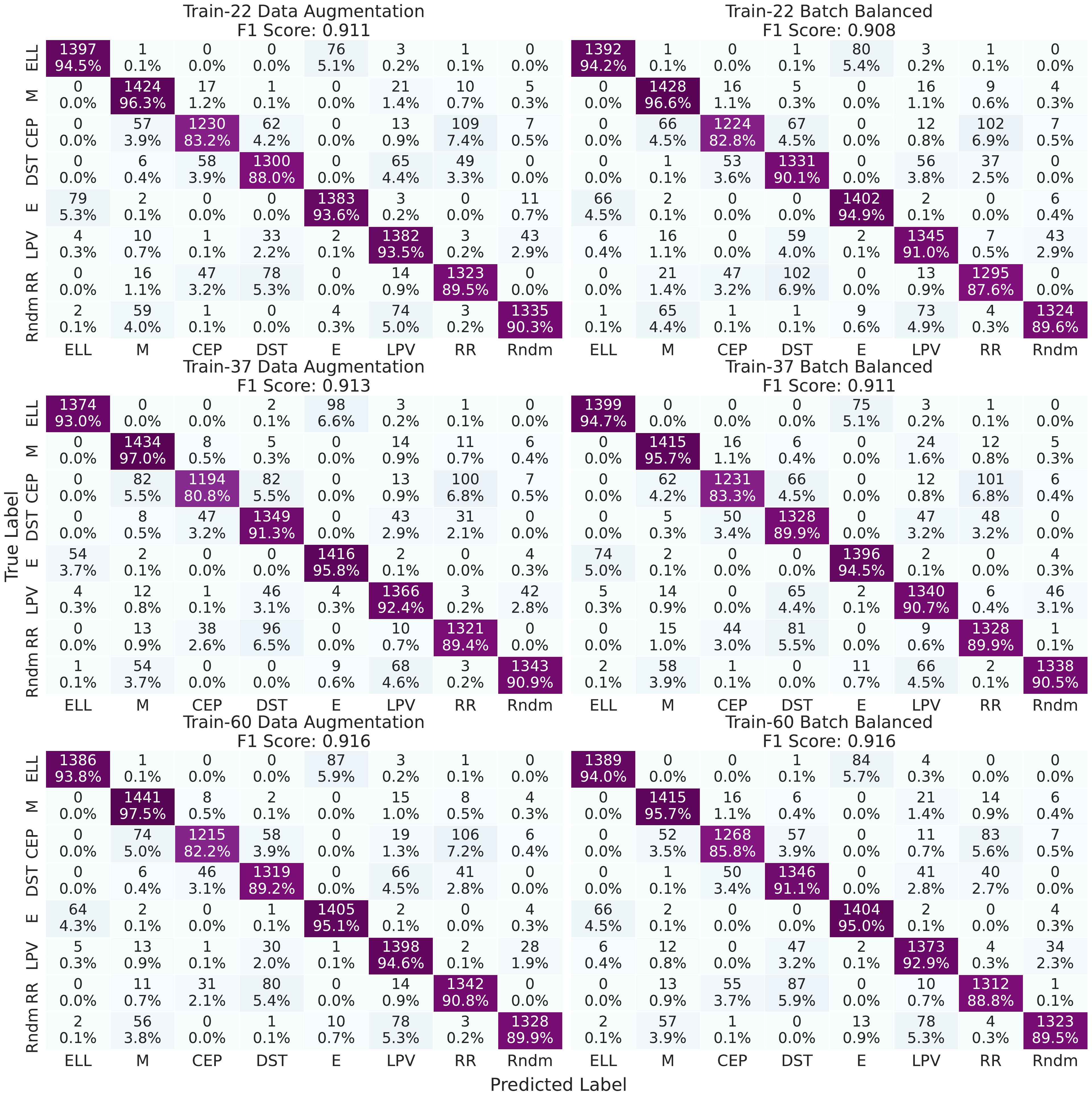}
  \end{center}

  \caption{\small Confusion matrix for the convolutional neural network models in differents training sets calculated in the same test set. The Y-axis is the true label obtained for OGLE, and the x-axis is the label predicted by the CNN. The boxes show the number of stars classified in each class and the percentage of the total sample.}
  \label{fig:CM}
\end{figure*}
The training process aims at establishing a model ($f$) that classifies the 2D histogram of the LC ($x$). We define the architecture of the CNN classifier as $f_{CNN}(x,\theta)$. During training, we adjust the free parameters ($\theta$) to create a CNN that effectively deal with the classification task: $f \approx f_{CNN}(x,\theta)$. \\
We employ categorical cross entropy as our loss function ($J$) for classification. This loss function measures the discrepancy between the predicted distribution of the model and the actual class distribution. Minimizing this function narrows the gap between the prediction of the model and the true classes, thereby improving the classifier performance. \\
We used Adaptive Moment Estimation \citep{2014arXiv1412.6980K} to minimize the loss function $J$. The parameters employed were $\eta = 0.0001$, $b_{1}=0.9$, $b_{2}=0.999$, and $\epsilon = 0.1$. $\eta$ is the learning rate and is the proportion of updating the parameters. The $b_{1}$ and $b_{2}$ are the exponential decay rate for the first and second moments, and the $\epsilon$ is a small constant for numerical stability (see \cite{2016arXiv160904747R} for an in-depth discussion on gradient descent optimization).\newline
We chose a batch size of 64 for the training sets and for the validation set. We trained the model for 1000 epochs, but stopped the training process when the algorithm showed indications of overfitting using the early stopping technique. This method actively monitors a designated quantity and ceases training when the quantity exhibits no improvement over a predefined number of epochs, termed 'patience'. For our purposes, we selected the validation loss as the quantity to monitor and established a patience threshold of 15 epochs. 

\section{Results}\label{sec:result}
\subsection{Training performance}
Figure \ref{fig:training} illustrates the training process for the different training sets using various balancing techniques (see Table \ref{table:Trainings_sets}). The column titles indicate the name of each training set. Across the three columns, the training process for each model is depicted. The first row represents the accuracy plotted against the epoch, while the second row shows the loss against the epoch. \newline
In Train-8 U, we present the training process for the balanced set using U. We achieved training and validation accuracies of $\sim 90\%$ and, $\sim 88\%$ respectively. The loss values for training (represented by the yellow line) began to drop slightly lower than those for validation (represented by the purple line). This pattern indicates an overfitting, where the model starts to learn the training set more effectively than the validation set. Given that these sets are independent, such overfitting can be interpreted as a halt in the improvement of the model, suggesting that further training may not be necessary. \newline
For Train-22, Train-37 and Train-60, we display two different training processes: the solid line represents DA, while the dashed line represents BB. Both balancing approaches yielded similar results. We noted slight overfitting for the DA method and a smoother training curve for the BB technique. However, the BB technique needs more epochs to converge than the DA method. \newline
\subsection{Metrics}
We use four well-known performance metrics to evaluate the classification model: Accuracy, Precision, Recall, and the F1 Score. For an individual class \textit{i}, these scores are defined as follows:
\begin{equation}
\text{Accuracy}_i = \frac{\text{TP}_i + \text{TN}_i}{\text{TP}_i + \text{TN}_i + \text{FP}_i + \text{FN}_i}
\end{equation}
\begin{equation}
\text{Precision}_i = \frac{\text{TP}_i}{\text{TP}_i + \text{FP}_i}
\end{equation}
\begin{equation}
\text{Recall}_i = \frac{\text{TP}_i}{\text{TP}_i + \text{FN}_i}
\end{equation}
\begin{equation}
\text{F1 Score}_i = 2 \times \frac{\text{Precision}_i \times \text{Recall}_i}{\text{Precision}_i + \text{Recall}_i}
\end{equation}

Where \( \text{TP}_i \) represent the number of true positives, \( \text{FP}_i \) false positives, \( \text{FN}_i \) false negatives, and \( \text{TN}_i \) true negatives per class. Multiclass metrics are calculated as the average of multiple binary classification problems, one for each label. \\
To obtain a comprehensive view of a classifier's performance, it is necessary to use more than one metric. While accuracy monitors the overall performance across all classes, it does not provide detailed information about the reliability or the model's precision. \newline
We used the seven CNN models, trained with different training sets, to obtain classification predictions for the LCs in the test set. The global performance is shown in Table \ref{table:metrics_diferent_trainings}. As we are using a balanced test set, we obtain similar values for accuracy, precision, recall, and F1 score. The Train-8 U serves as our baseline, with an F1 score of $0.891$. The inclusion of more data improves the model’s performance across various training sets. Train-22 DA and Train-22 BB show that DA yields better results, enhancing the model's performance. Train-37 DA and Train-37 BB result in a slight improvement, with BB providing similar performance to DA. Train-60 continues this trend, displaying similar outcomes for both BB and DA techniques.
\begin{table}[ht]
  \begin{center}
    \caption{Performance metrics for different models.}
    \label{table:metrics_diferent_trainings}
    \begin{tabular}{ccccc} 
   \hline
Training Set  & Accuracy & Precision & Recall & F1  \\
&  &  \\
\hline
\hline
Train-8 U & 0.890 & 0.890 & 0.890 & 0.891 \\
Train-22 DA & 0.911 & 0.911 & 0.911 & 0.912 \\
Train-22 BB & 0.908 & 0.908 & 0.908 & 0.909 \\
Train-37 DA & 0.913 & 0.913 & 0.913 & 0.915 \\
Train-37 BB & 0.911 & 0.911 & 0.911 & 0.912 \\
Train-60 DA & 0.916 & 0.916 & 0.916 & 0.918 \\
Train-60 BB & 0.916 & 0.916 & 0.916 & 0.917 \\
\hline
  \end{tabular}
  \end{center}
  \tablefoot{trained on various balanced training sets and evaluated on the same balanced test set.}
\end{table}
\subsection{Class performance}
The results for each class are presented in the confusion matrices in Fig. \ref{fig:CNN_Undersampling} and Fig.\ref{fig:CM}. The title of each figure displays the name of the training set, the balancing technique used, and the F1 score of the model. First, we present the results in our baseline Train-8 U. Subsequently, we display the outcomes for the different balancing techniques.\newline
The most frequent misclassifications occur among pulsating stars (RR, M, DST, and CEP), which are in different evolutionary states. The amplitude and period are important parameters for distinguishing between pulsating stars. However, the algorithm does not directly access the values of period and amplitude. Instead, it indirectly accesses the periods within the pixels of the LCs, which appear much larger in objects with shorter periods. But, due to the different cadences, number of observations, and baselines, we obtain a certain degree of imprecision.\newline
The other misclassifications occur among classes that denote binary star systems (E, ELL). The LC of an E results from the eclipses of the companion star. Assuming the binary does not undergo an eclipse and is close ($a \lesssim 15\text{R}_\odot$), the most pronounced feature in the LC emerges from ellipsoidal modulation due to tidal deformation \citep{2023MNRAS.522...29G}. This LC are typically close to sinusoidal with two equal maxima and minima in the phase \citep{2014AcA....64..293P}. Therefore, binary systems can exhibit a combination of both eclipsing and ellipsoidal effects. This overlap of characteristics presents challenges, making accurate classification difficult. \newline
Finally, the classification error between LPV and DST stars is not immediately apparent. DST stars are faint main sequence stars with I-band magnitudes that range from 19 to 21 magnitudes in the Large Magellanic Cloud. This puts them close to the detection limit of OGLE \citep{2010AcA....60....1P}. As a result, the noisy LCs of DST stars can resemble those of LPV. \newline

\subsection{Balancing strategies}
Figure \ref{fig:CNN_Undersampling} shows the main results with the different balancing techniques. Only specific classes require balancing, CEP in Train-22, both CEP and ELL in Train-37, and CEP, ELL, DST in Train-60. CEP does not show improvement from Train-8 U to Train-60 using the DA technique, but experiences improved performance with Train-60 BB. Similarly, ELL shows only slight improvement from Train-22 to Train-37 using DA, reaching its best performance with Train-37 BB. DST follows a similar pattern, with performance decreasing from Train-37 to Train-60 DA, and the most successful results observed with Train-60 BB. These patterns indicate that the BB technique provides better performance improvement for the classes requiring balancing. Additionally, the overall good performance of DA is due to the use of more data in other classes than to an improvement in the balanced class.\newline
Out of the seven models, Train-60 classes shows the best overall performance, and the BB give the best results for the balanced classes. We use this CNN to report the metrics in Table \ref{precision_recall}. The algorithm performs with an F1 score greater than 0.88 for all classes. ELL and E have similar values for precision, recall, and F1 score. We expect a similar distribution of False Positives and False Negatives for these misclassifications. The CEP, RR, and Random period classes have higher precision than recall. We expect more False Negatives than False Positives, meaning the algorithm is more likely to miss these classes and not identify all the stars, but the classifications made are precise. The M, DST, and LPV classes have higher recall than precision. We expect more False Positives than False Negatives, suggesting we can identify the majority of the stars in these classes, but may also find contaminants. The DST class has the lowest precision, so we expect it to be the most contaminated predicted class. \newline

\begin{table}[htpb]
  \begin{center}
    \caption{Performance metrics obtained for the train-60 BB model}
    \label{precision_recall}
    \begin{tabular}{cccc} 
      \hline
      Class & Precision & Recall & F1-score \\
      \hline
      \hline
      ELL & 0.949 & 0.940 & 0.945 \\
      M & 0.912 & 0.957 & 0.934 \\
      CEP & 0.912 & 0.858 & 0.884 \\
      DST & 0.872 & 0.911 & 0.891 \\
      E & 0.934 & 0.950 & 0.942 \\
      LPV & 0.892 & 0.929 & 0.910 \\
      RR & 0.900 & 0.888 & 0.894 \\
      Rndm & 0.962 & 0.895 & 0.927 \\
      \hline
    \end{tabular}
  \end{center}
\end{table}
\subsection{Convolutional neural network with additional Information}
\begin{figure}
  \begin{center}
    \includegraphics[scale=0.5]{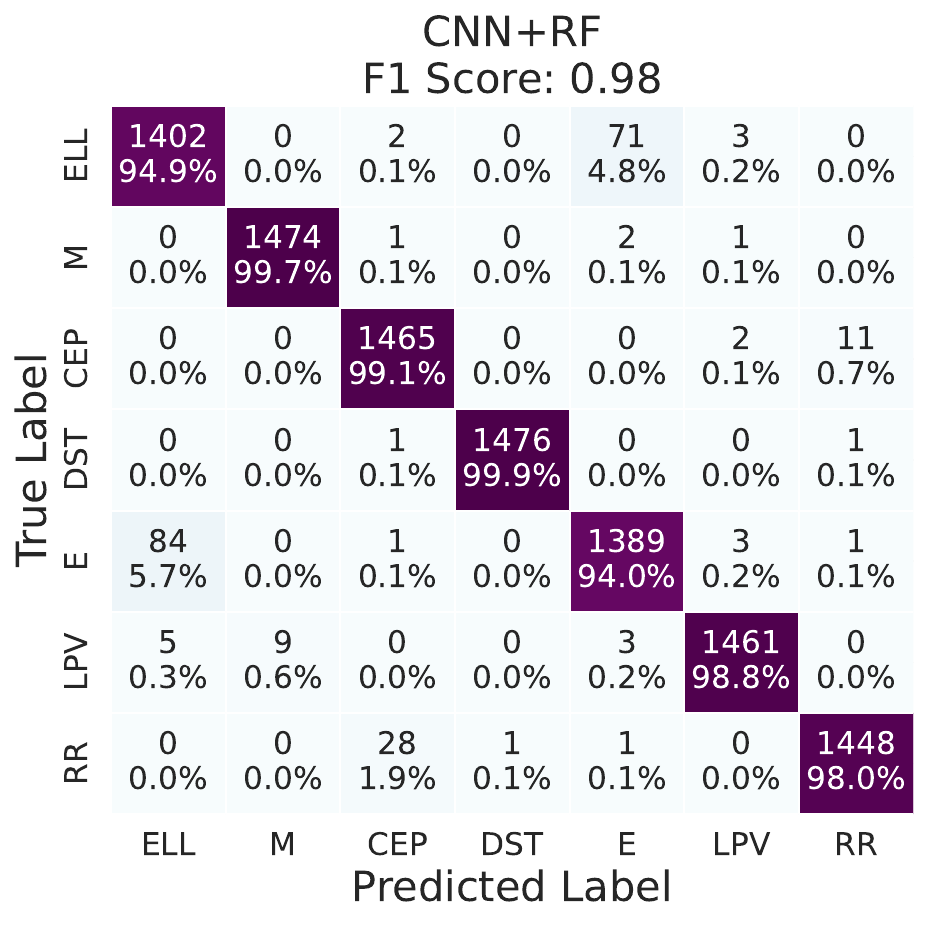}
  \end{center}
  \caption{\small Confusion matrix for the RF classifier using the output of the CNN together with amplitude and period. The algorithm was trained using the output of the train-37. We used the same Training, Validation, and Testing sets as in the Convolutional Neural Network.}
  \label{fig:CNN_RF}
\end{figure}

We compose a two-step classifier that combines the CNN with a RF that uses our convolutional model to enhance LC classification. Our CNN extracts the visual information from LCs. Additionally, we aim at integrating astrophysical knowledge, especially period and amplitude, to complement this visual insight. \newline
We present an implementation aligned with RF due to its strong performance. RF is a supervised machine learning algorithm that functions as an ensemble of decision trees \citep{breiman2001random}. Each decision tree defines a potential decision path based on a subsample of the features in the training set. The final decision is the average of the decisions of each tree. We configure a RF using the default parameters of \texttt{scikit-learn} \citep{scikit-learn} and with 10 trees. \newline
We employ the Train-60 BB model to predict the variability classes for the Train-8 U, validation, and test sets, initially generating eight columns of data. These columns represent the probabilities assigned by the Train-60 BB model to each of the eight variability classes. We aggregate the period and amplitude for each star, generating tabular data comprising ten features. These features are the input for the RF algorithm. Unlike the CNN, the RF does not require preprocessing steps such as normalization, so we do not normalize our features. We maintain consistency between the training and testing sets. In our RF analysis, we combine the validation set with the training set, as the RF method does not require a separate validation set.\newline
Figure \ref{fig:CNN_RF} presents the confusion matrix resulting from the RF algorithm. The combination of classifications obtained via 2D histogram phased LCs plus a few attributes enhances performance, as evidenced by improvements in the confusion matrix and the F1 score. This simple methodology demonstrates the flexibility of image classification and its synergy with other approaches. The combination of these methods enables high-precision recovery of all OGLE classes in this study. \newline
We emphasize that we do not directly use the period and amplitude information as inputs for the CNN. The period is not explicitly included, but in a way, it is through the intensity of the pixels ``in'' the LC. Our goal is to develop a robust method for classifying variable stars. The main challenges include the effects of observational capabilities on amplitude variability detection and potential misclassification of periods.

\section{Discussion}\label{sec:discussion}
In this work we have presented an efficient methodology that combines two classifiers to codify objects in different variability classes. In the following we will discuss the astrophysical and computational performances of our methodology separately.\newline

\subsection{Comparison with previous works}
Our research focuses on the morphological classification of variable stars using image-based classifications with phasefolded LCs. This approach is fundamentally different than other machine learning methods applied to the sequential representation of the same or similar data, for example applying Recurrent Neural Networks (RNN; \citeauthor{2020MNRAS.493.2981B} \citeyear{2020MNRAS.493.2981B}), Neural Networks (NN; \citeauthor{2016A&A...587A..18K}\citeyear{2016A&A...587A..18K}), Random Forests (RF; \citeauthor{2021AJ....161..141S} \citeyear{2021AJ....161..141S}), or more recently, Transformers \citep{2023A&A...670A..54D}. In order to ease direct comparison of the results, in this section we restrict ourselves to studies based on similar taxonomy and OGLE data. However, we do compare with methodologies based not only on image representation but also ``classical'' sequential representation.\newline

\subsubsection{Image based representations}
\cite{2020ApJ...897L..12S} (S20), explored the use CNNs for classifying image representation phasefolded LCs from OGLE. In \cite{2022ApJ...938...37S} (S22), they broaden their methodology by incorporating a multi-input neural network into the CNN.\newline

\begin{table}[htpb]
  \begin{center}
    \caption{F1 Score for the Variability classes Common to S20.}
    \label{table:comparasion}
    \begin{tabular}{cccc} 
      \hline
      Class & CNN this work & S20 OGLE III & S20 OGLE IV \\
      \hline
      \hline
      ACep & 0.884 (CEP) & 0.879 & 0.835 \\
      T2Cep & 0.884 (CEP)     &  0.872  &  0.870 \\     
      DST & 0.891 & 0.944 & -- \\
      E & 0.942 & 0.989 & 0.984 \\
      RR & 0.894 & 0.810 & 0.849 \\
      \hline
    \end{tabular}
  \end{center}
  \tablefoot{We adopt the F1 score for CEP as the average of the F1 scores for Acep and T2Cep in S20.}
\end{table}
S20 presents and analyses a LMC dataset with 26 121 E, 24 904 RR, 2696 DST, 83 Anomalous CEP, and 203 Type 2 CEP. Their test set comprises 3750 augmented images for each variability type in OGLE III and 2500 augmented images for each type in OGLE IV. We compared the results of S20 with our CNN results, specifically for the variability classes that are common to both studies. Table \ref{table:comparasion} shows the comparison between the F1 score from S20 and this work. In general, their results were similar than ours; however, the creation of artificial LCs, sampled purely accounting for photometric errors (as is the strategy for data augmentation employed in S20), does not guarantee the independence of instances in the training and testing sets. Consequently, performance metrics might be artificially biased in a positive manner. \newline
S22 expands on the work of S20 by using LCs from the Magellanic Clouds, Galactic Bulge, and Galactic Disk. In addition, the CNN results are combined with period information to classify the six main variability classes. S22 do not provide a tabular from of the metrics achieved, therefore, we compared the confusion matrices between they multiple input neural network and our combination of CNN and RF (CNN+RF). We interpret the better performance of our model for the CEP and RR classes as a result of incorporating amplitudes as features and we do not use T2 CEP and anomalous CEP. We obtained similar results for DST; this can be mostly for the period values that are smaller than the others pulsational periods and can be easier to distinguish. Finally, we have a worse performance for E, this can be explained by the fact that they are not using ELL. Overall, our results are better for the main classes, but, the inclusion of DA in the test set is a fundamental difference that makes it difficult to compare algorithms. 
\subsubsection{Time Series Representation}
\cite{2019MNRAS.482.5078A} presents a Deep Learning method based on Convolutional units to classify variable stars across multiple surveys, including OGLE, VVV \citep{2010NewA...15..433M}, and COROT \citep{2003AdSpR..31..345B,2003A&A...405.1137B}. The model inputs are the differences in time and magnitude of the LC. It is trained with 8000 stars per class and survey, with a maximum of 500 observations. Additionally, a RF is trained to compare with the CNN results, extracting 59 features for each survey using the FATS library \citep{2015arXiv150600010N}. Table \ref{table:ogle_aguirre} compares the accuracy reported in \cite{2019MNRAS.482.5078A} with our results. The combined CNN+RF model achieves better results for RR and CEP classes. We obtained marginally better accuracy for three out of the four classes in common. For E class, the lower performance is attributed to misclassification with the ELL class.

\begin{table}[htpb]
  \begin{center}
    \caption{Accuracy of OGLE Observations as presented in \cite{2019MNRAS.482.5078A} compared to the results of this work.}
    \label{table:ogle_aguirre}
    \begin{tabular}{cccc} 
      \hline
      Class &CNN+RF& Aguirre CNN & Aguirre RF \\
      \hline
      \hline
      E & 0.942 & 0.98 ± 0.01 & 0.97 ± 0.01 \\
     LPV & 0.991 & 0.99 ± 0.00 & 0.97 ± 0.01 \\
      RR & 0.990 & 0.94 ± 0.01 & 0.97 ± 0.00 \\
      CEP & 0.983 & 0.90 ± 0.03 & 0.93 ± 0.01 \\
      \hline
    \end{tabular}
  \end{center}
\end{table}
\cite{2020MNRAS.493.2981B} presents a classification model based on RNNs, which uses the differences in time and magnitude as input. This method is tested across three different surveys: OGLE-III, \textit{Gaia} DR2 \citep{2018A&A...616A...1G}, and WISE \citep{2010AJ....140.1868W}. The OGLE dataset incorporates a total number of 393103 stars. For comparison, a RF with 1000 trees, utilizing 59 single-band features from the FATS Library, is employed. Table \ref{table:metrics_rnn_rf} shows the F1 scores for the classes, compared with our work, where better results are achieved compared to the RNN and similar outcomes to the RF. However, the processing time using FATS for the RF is $\sim$ 7 days. Additionally, \cite{2020MNRAS.493.2981B} highlights that the OGLE LCs are biased, as they were selected from a feature-based classification, which favors these models over others.
\begin{table}[htpb]
  \begin{center}
    \caption{F1 Score of OGLE Observations as presented in \cite{2020MNRAS.493.2981B} compared to the results of this work.}
    \label{table:metrics_rnn_rf}
    \begin{tabular}{cccc} 
      \hline
      Class&CNN+RF & RNN F1-score & RF F1 score \\
      \hline
      \hline
      CEP &0.99 & 0.69 & 0.97 \\
      RR  & 0.99 &0.91 & 0.99 \\
      DST & 1.00 & 0.72& 0.95 \\
      E   & 0.95 & 0.94&0.98 \\
      LPV & 0.99 & 0.99& 1.00 \\
      \hline
    \end{tabular}
  \end{center}
\end{table}
\cite{2021MNRAS.505..515Z} introduce Cyclic-Permutation Invariant Neural Networks designed to be invariant to phase shifts of period-folded periodic LCs. They showcase the implementation of this neural network type using 1d Residual Neural Networks (ResNets; \citeauthor{he2016deep} 2016) and Temporal Convolutional Networks (TCN; \citeauthor{lea2016temporal} 2016), with the model input being the difference in phase and normalized amplitude. Utilizing data from OGLE III, they segment 163356 stars into chunks of fixed length, resulting in 540457 fixed-length LCs, and emphasize that using sequences of varying lengths can degrade accuracy. We achieve comparable results with both networks; however, we observe higher performance for E class.

The experimental design in this study differs from previous works in two main aspects: base datasets and taxonomy. While the other studies that we are comparing to used only OGLE III data (with the exception of a test set in S20), we combined all available OGLE III and OGLE IV data for objects with classes within our taxonomy. Regarding the latter, the taxonomy varies across studies. Therefore, including different subtypes of variability affect the comparisons. For example, the performance of our method for eclipsing binaries is lower than those in some of the other studies (S20 and \cite{2019MNRAS.482.5078A}), primarily due to the difficulty in differentiating eclipsing binaries from ellipsoidal variables (joint in a single class in the aforementioned studies).

The performance in terms of ``purity'' and other metrics is hard to assess since it can be affected by the different taxonomies (and, probably to a lesser extent, the base dataset themselves). However, the performance in terms of computing time is indeed comparable because we have a similar length of time series in OGLE III and OGLE IV (with a K-S test $\sim$ 0.2), and these are directly proportional to the feature extraction time. A more direct comparison should involve using the same datasets and taxonomy, however this is not currently possible. The data links in \cite{2020MNRAS.493.2981B} and \cite{2019MNRAS.482.5078A} are not available, and the S20 data is not publicly accessible. In S22 and \cite{2021MNRAS.505..515Z} they provided GitHub reference; But we could not find an OGLE list to identify the stars used for training and validation, making it impossible to identify independent stars for testing.

\begin{table}[htpb]
  \begin{center}
    \caption{Accuracy Score of OGLE observations as presented in \cite{2021MNRAS.505..515Z} for iTCN and IResNet Models compared to the results of this work.}
    \label{table:performance_itcn_iresnet}
    \begin{tabular}{cccc} 
      \hline
      Class &CNN+RF & iTCN (\%) & IResNet (\%) \\
      \hline
      \hline
      Cep & 98.31 & 98.3 ± 0.3 & 98.4 ± 0.7 \\
      RRab & 98.99 (RR) & 99.7 ± 0.1 & 99.7 ± 0.4 \\
      RRc & 98.99 (RR) & 99.0 ± 0.2 & 99.1 ± 0.1 \\
      Dsct &99.73 & 97.6 ± 0.8 & 97.8 ± 0.6 \\
      EC & 94.18 (E) & 87.9 ± 0.9 & 87.8 ± 0.7 \\
      ED & 94.18 (E) & 95.0 ± 0.3 & 94.8 ± 0.4 \\
      ESD & 94.18 (E) & 68.7 ± 1.0 & 70.7 ± 0.9 \\
      Mira & 99.73 & 97.1 ± 0.6 & 96.8 ± 0.3 \\
      SRV & 99.12 (LPV) & 96.0 ± 0.4 & 95.9 ± 0.2 \\
      OSARG& 99.12 (LPV) & 93.2 ± 0.4 & 93.4 ± 0.2 \\
      \hline
    \end{tabular}
  \end{center}
\end{table}
\subsection{Computational resources}
\begin{figure}
  \begin{center}
    \includegraphics[scale=0.6]{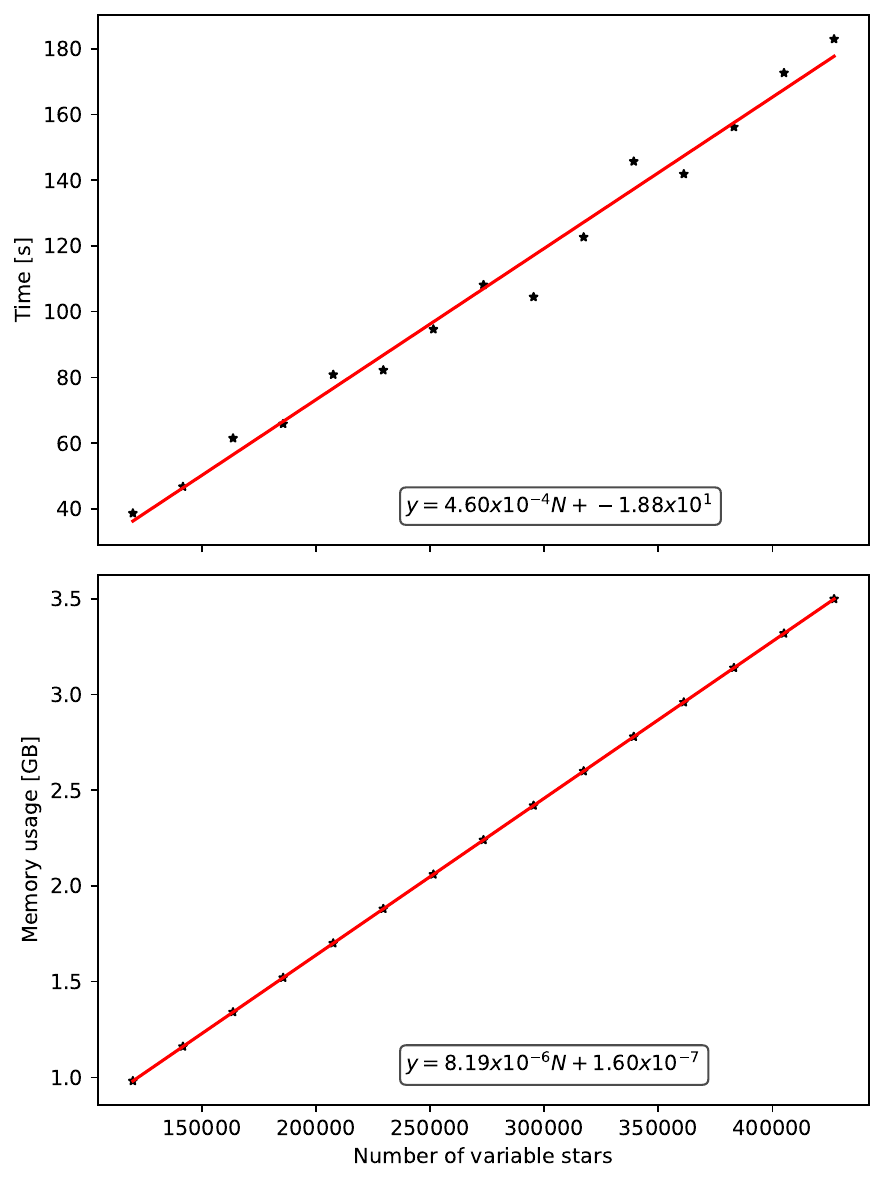}
  \end{center}
  \caption{Computational resources required to implement the methodology of 2D histogram-based convolutional neural network for variable stars. The Figure presents the relationship between prediction time and memory usage for N variable stars represented in 2D histograms.}
  \label{fig:Resources}
\end{figure}
As previously mentioned, our goal with this project is two-fold: to produce an accurate classifier, but also an efficient one that does not involve heavy computational resources.\newline
Our input image size is 32x32 because it allows the analysis of more LCs with less memory. If we double the size of the 2D histograms to 64x64, we could use a deeper neural network and therefore may achieve better classification results (see for example \cite{2019arXiv190511946T}). However, the double size of the images quadruples the memory space they used. Therefore, it is necessary to find a trade-off between the size of the images, the complexity of the network, the performance of the model and the computational resources. We decided to give priority to the speed and efficiency maintaining a competitive performance of the algorithm. \newline
We present two time phases: the Train Time, which includes creating histograms and train the CNN, and the Prediction Time. The Train Time is carried out on a computer with 64 GB of memory, an Intel Core i9, and an Nvidia GeForce RTX 4060 GPU. For prediction, we utilize a computer with 8 GB of memory, an Intel Core i5, and a 256 GB SSD.
\subsubsection{Training time}
Table \ref{table:time_extraction_training} shows the time required for feature extraction and model training for a sample of 90,928 stars. The feature extraction process for the CNN is the creation of the 2D histograms, with 33\% of the time dedicated to opening CSV data, 37\% to phase folding and sigma clipping, and 30\% to the generation of the 2D histograms. The feature extraction time for the this work RF is approximately equivalent to the CNN's time to generate the probabilities, which then serve as inputs for the RF. The RF model's training time is brief as we are utilizing only 10 trees. For comparison, we reference the time reported by \cite{2019MNRAS.482.5078A}. which used a complete sample of 51,951 stars. It is important to note that from this sample they use a subsample as training set, alongside the use of DA, allowing for a maximum of 5 augmented stars per LC. The specifics of the training set size, however, remain undefined.
\begin{table}[htpb]
  \begin{center}
    \caption{Approximate Time for feature extraction and training of algorithms.}
    \label{table:time_extraction_training}
    \begin{tabular}{cccc} 
      \hline
      Method & \makecell{Extraction of \\ Features} & \makecell{Training \\ Algorithm} & Total Time \\
      \hline
      \hline
      This work CNN & 50 s & 13.9 min & 14.73 min \\
      This work RF & 5.44 s & 1.48 s & 6.92 s \\
      Aguirre RF & 11.5 days & 36 min & 11.52 days \\
      Aguirre CNN & 30 min & 50 min & 1.33 hrs \\
      \hline
    \end{tabular}
  \end{center}
  \tablefoot{We analyze a sample of 90 928 stars using 24 cores for parallel feature extraction and an RTX 4060 GPU for training. For comparison, we reference the time calculated by \cite{2019MNRAS.482.5078A}, who utilized 6 CPUs for parallel feature extraction and a GeForce GTX 1080 Ti GPU, For a original sample of 51 951 stars.}
\end{table}
In summary, our two-layered model is able to achieve results comparable to previous works. However, there are two main advantages to our approach: On the one hand, the identification of miscalculated periods (a ``class'' commonly ignored in the literature), and on the other, the computational time efficiency gain. Regarding the former, the accurate and precise identification of periods for all variability classes presents challenges, as it is dependent on both the survey characteristics and the variability classes themselves \citep{2013MNRAS.434.3423G}. Therefore, the ``spurious class'' can be used as a diagnostic on the reliability of reported periods. 

The time efficiency is achieved by minimizing the number of tabular features that need to be calculated, avoiding the use of time-consuming algorithms for feature extraction, such as FATS (see Table \ref{table:time_extraction_training}). One example of a state-of-the-art approach that uses feature extraction is the Automatic Learning for the Rapid Classification of Events (ALeRCE, \cite{2021AJ....161..242F}). It processes alerts from ZTF, which contain photometry in g and r bands. It employs a Balanced Random Forest \citep{chen2004using} with 500 trees and 152 recent and optimized feature extraction packages \citep{2021AJ....161..141S}. A fair comparison with their tools in production is not possible because the data types are different (they process ZTF Avro alert files 4\footnote{\href{https://zwickytransientfacility.github.io/ztf-avro-alert/}{https://zwickytransientfacility.github.io/ztf-avro-alert/}}). Furthermore, the LC classifier of ALeRCE tackles a more general problem than the one addressed here. But, since their architecture and procedures are public, we conducted a comparison experiment: We selected randomly balanced samples of stars with different amount of epochs of observations (70 stars per bin on the numbers of epochs as shown in Table \ref{table:FE_CNN_RF}). For the performance comparison, we focused on 61 single-band features from the ALeRCE pipeline (61-AF) and timed the feature extraction plus classification flow using a 500-tree forest. These figures (computing time) were contrasted with those from our approach: feature extraction and classification with 10 trees. Table \ref{table:FE_CNN_RF} shows the results of the comparison. Both experiments were conducted without parallelization and in the same personal laptop. As can be seen (and somewhat expected intuitively), the feature extraction time increases significantly with the number of observations, whereas our method maintains a constant time regardless of the number of observations. This experiment confirms the intuitive idea that our methodology computational requirements scales (in the sense of actually not scaling) nicely with density / size of the lightcurves and hence it is a good alternative to traditional methods. However, we must emphasize in the methodologies as future facilities will provide both, dense and sparse data sets and will benefit from focused / smaller and wider taxonomies.

\begin{table}[ht]
  \begin{center}
    \caption{Classification time for 70 stars per number of observation bin.}
    \label{table:FE_CNN_RF}
    \begin{tabular}{lcc} 
    \hline
    $n_{obs}$ & 61-AF+RF [s]& CNN+RF [s] \\
    \hline
    \hline
    (60, 100] & 0.142 $\pm$ 0.042 & 0.034 $\pm$ 0.001 \\
    (100, 500] & 0.486 $\pm$ 0.284 & 0.034 $\pm$ 0.002 \\
    (500, 800] & 1.337 $\pm$ 0.385 & 0.034 $\pm$ 0.001 \\
    (800, 1500] & 2.696 $\pm$ 0.579 & 0.034 $\pm$ 0.001 \\
    (1500, 2000] & 4.533 $\pm$ 1.262 & 0.034 $\pm$ 0.001 \\
    \hline
    \end{tabular}
  \end{center}
\tablefoot{We show the time for 61 single-band features from the ALeRCE pipeline and compare it with the classification time of CNN+RF.}
\end{table}

In a forthcoming paper, we aim to investigate the synergy between our method and various periodograms to identify the optimal combination of methods for accurately recovering the variability class with a reliable period.
\subsubsection{Predicting time}
We tested the algorithm on a 8GB of memory, Intel Core i5, and a 256 GB SSD. In Fig. \ref{fig:Resources} we show the time and memory used to classify different samples of stars. \newline
We can classify and represent half a million stars in 3 minutes and using 3.5 GB of memory, respectively. We do not consider the time required to create a 2D histogram. However, we have made the complete OGLE catalog used in this work publicly available, along with the corresponding 2D histogram for the OGLE data, on GitHub\footnote{\href{https://github.com/Monsalves-Gonzalez-N/Paper_OGLE}{https://github.com/Monsalves-Gonzalez-N/Paper\_OGLE}}. \newline
These results are promising and competitive in terms of speed and accuracy. As \citep{2023arXiv230201436C} indicates, the \textit{Vera Rubin} Observatory is expected to detect up to $10^{8}$ variable stars. Our algorithm is potentially able to process this volume of data in approximately 13 hours, but it may not be feasible to handle such a quantity on a standard computer. Therefore, further research is necessary to optimize data representation and analysis for more accessible processing. This big data scenario marks a significant paradigm shift in scientific research methodology. Consequently, there is a pressing need to develop tools that are accessible to the broader community of astronomers, particularly those with limited access to high computational resources, with the ultimate goal of democratizing science.\newline

\section{Summary and conclusions}\label{sec:summary}
In this study, we have introduced a methodology to classify variable stars based on the morphology of their LCs. We present a 2D histogram, sized 32x32, to depict LCs as images. We introduce a Convolutional Neural, consisting of convolutional layers, max pooling, and dense layers.\newline
We select our variable stars data from the OGLE. We choose variability classes with a reasonable number of examples per class. Using these stars, we select eight classes: RR, CEP, LPV, M, ELL, DST, E, and an additional artificial class representing LCs with misclassified periods.\newline
We applied three distinct approaches to manage the unbalanced dataset: DA, BB, and U. BB emerged as the most effective, achieving an F1 score of $\sim$0.92. In contrast, DA led to less favorable outcomes in the augmented classes compared to those obtained with U. Our DA strategy includes phase shifts, random reductions in the number of LC observations, and magnitude shifts within error margins. The suboptimal performance of DA suggests that these techniques may not reliably provide novel and beneficial information to the algorithm.\newline
We achieve a limit in the image classification of $\sim 92\%$. We justify this value because it is challenging to differentiate pulsating stars without period and amplitude values. Thus, we test a two step algorithm with a RF that incorporates the neural network output along with period and amplitude values. With this adjustment, the convolutional network identifies almost all variability classes, except for distinguishing between Ellipsoidal and Eclipsing variations, where a $5\%$ discrepancy remains.\newline
Finally, we outline the resources necessary for implementing our proposed methodology, emphasizing the efficiency of our approach. Our method demonstrates notable efficiency: opening histograms for variable stars requires approximately 4 gigabytes of memory for half a million variable stars. Moreover, the classification process takes around 180 seconds for this volume of data on a standard computer. On a better computer, we are capable of creating approximately 76 images per second per core and a prediction time of $\sim$ 60$\, \mu\text{s}$ per star. Such efficiency is noteworthy and contributes to addressing the challenges of big data in astronomy.

\begin{acknowledgements}
We acknowledge support from the National Agency for Research and Development (ANID) through the Scholarship Program DOCTORADO BECAS CHILE/2021 - 21211323. Additionally, A.B acknowledges support from the Deutsche Forschungsgemeinschaft (DFG, German Research Foundation) under Germany's Excellence Strategy – EXC 2094 – 390783311. We offer our sincere thanks to the FIULS 2030 project (18ENI2-104235 - CORFO) for providing essential computing resources via the SynergyGrid Server. The provision of GPU time was partially supported by the ANID FONDECYT Regular grant number 1211370, with PI: Gómez. The referenced service for OGLE filter transmissivity curves is referenced by \cite{2024arXiv240603310R}
\end{acknowledgements}

%
%
\bibliographystyle{aa}
\bibliography{bibliography.bib}
\appendix
\onecolumn
\section{}
\begin{table}[htpb]
  \caption{Different works used in this paper for each variability class.}
  \label{table:references_for_classes}
  \centering
  \begin{tabular}{cc} 
    \hline
    variability class & Reference \\ 
\hline
 ELL & \makecell{\citep{2016AcA....66..405S}} \\
 M  & \makecell{\citep{2009AcA....59..239S, 2011AcA....61..217S, 2013AcA....63..379P, 2013AcA....63...21S} \\ \citep{2022ApJS..260...46I}} \\ 
 CEP  & \makecell{\citep{2008AcA....58..163S, 2010AcA....60...17S, 2011AcA....61..285S, 2013AcA....63..379P} \\ \citep{2015AcA....65..297S, 2018AcA....68..315U, 2017AcA....67..297S, 2020AcA....70..101S}} \\
 DST  & \makecell{\citep{2010AcA....60....1P, 2013AcA....63..379P, 2020AcA....70..241P, 2022AcA....72..245S} \\ \citep{2023AcA....73..105S}} \\
 E  & \makecell{\citep{2011AcA....61..103G,2013AcA....63..115P, 2013AcA....63..323P,  2016AcA....66..405S} \\ \citep{2016AcA....66..421P}} \\
LPV  & \makecell{\citep{2009AcA....59..239S, 2011AcA....61..217S, 2013AcA....63..379P, 2013AcA....63...21S}} \\
RR  & \makecell{\citep{2009AcA....59....1S, 2010AcA....60..165S, 2011AcA....61....1S, 2013AcA....63..379P} \\ \citep{2014AcA....64..177S,
2016AcA....66..131S, 2019AcA....69..321S}}\\ 
\hline
\end{tabular}
\end{table}

\begin{table*}[htpb]
  \begin{center}
    \caption{Selected architecture and hyperparameters.}
    \label{table:hyper_parameters}
    \begin{tabular}{cc}
      \hline
      Layer Type & Layer Description \\ 
      \hline
      Input & Size: 32x32x1\\
      Convolutional & No. of Filters: 16, Filter Size: 3, Activation: Relu, Padding: Same \\
      Convolutional & No. of Filters: 16, Filter Size: 3, Activation: Relu, Padding: Same \\
      Max-Pooling & Kernel Size: 2 \\
      Convolutional & No. of Filters: 32, Filter Size: 3, Activation: Relu, Padding: Same \\
      Convolutional & No. of Filters: 32, Filter Size: 3, Activation: Relu, Padding: Same \\
      Max-Pooling & Kernel Size: 2 \\
      Fully Connected & No. of Neurons: 1024, Activation: ReLu\\
      Dropout & Dropout Probability: 30\% \\
      Fully Connected & No. of Neurons: 512, Activation: ReLu\\
      Dropout & Dropout Probability: 30\% \\
      Fully Connected & No. of Neurons: 8, Activation: Softmax\\
      \hline
    \end{tabular}
  \end{center}
\end{table*}

\end{document}